\documentclass[%
twocolumn,
superscriptaddress,
 amsmath,amssymb,
 aps,
]{revtex4-2}

\usepackage{graphicx}% Include figure files
\usepackage{dcolumn}% Align table columns on decimal point
\usepackage{bm}% bold math
\usepackage{hyperref}% add hypertext capabilities
\usepackage{physics}
\usepackage{mathtools}
\usepackage{mathrsfs}
\usepackage{siunitx}
\usepackage{lineno}

\usepackage{xcolor}
\definecolor{amber}{rgb}{1,0.49,0}
\definecolor{darkgreen}{rgb}{0,0.55,0}

\DeclareFontFamily{U}{mathx}{\hyphenchar\font45}
\DeclareFontShape{U}{mathx}{m}{n}{<-> mathx10}{}
\DeclareSymbolFont{mathx}{U}{mathx}{m}{n}
\DeclareMathAccent{\widebar}{0}{mathx}{"73}

\usepackage[normalem]{ulem}
\definecolor{tangerine}{rgb}{0.944,0.522,0}
\definecolor{verde}{rgb}{0.,0.6,0}
\definecolor{rosso}{rgb}{0.9,0.0,0.2}
\definecolor{magenta}{rgb}{0.9,0.2,0.9}

\newif\ifhighlight

\newcommand{\highlight}{\highlighttrue}
\highlight
\newcommand{\editor}[2]{%
  \expandafter\newcommand\csname #1note\endcsname[1]{%
    \textcolor{#2}{(\textbf{#1:} ##1)}}%
  \expandafter\newcommand\csname #1\endcsname[1]{%
    \ifhighlight\textcolor{#2}{##1} \else ##1\fi}%
  \expandafter\newcommand\csname #1cancel\endcsname[1]{%
    \ifhighlight\textcolor{#2}{\sout{##1}}\fi}%
  \expandafter\newcommand\csname #1change\endcsname[2]{%
    \ifhighlight\textcolor{#2}{\sout{##1} ##2}\else ##2\fi}%
  \newenvironment{#1text}{\ifhighlight\color{#2}\fi}{\color{black}}
}

\editor{SB}{tangerine}
\editor{PP}{verde}
\editor{resub}{cyan}
\editor{AF}{blue}
\editor{ED}{purple}
\editor{CAVEAT}{red}

\newcommand{\asic}{a\mathrm{SiC}}
\newcommand{\asio}{a\mathrm{SiO_2}}
\newcommand{\asi}{a\mathrm{Si}}

\usepackage{todonotes}
\usepackage{orcidlink}

\begin{document}

\title{Unearthing the foundational role of anharmonicity in heat transport in glasses}

\author{Alfredo Fiorentino\,\orcidlink{0000-0002-3048-5534}}
\affiliation{%
 SISSA---Scuola Internazionale Superiore di Studi Avanzati, Trieste
}%
\author{Enrico Drigo\,\orcidlink{0000-0002-1797-2987}}
\affiliation{%
 SISSA---Scuola Internazionale Superiore di Studi Avanzati, Trieste
}
\author{Stefano Baroni\,\orcidlink{0000-0002-3508-6663}} %\email{baroni@sissa.it}
\affiliation{%
 SISSA---Scuola Internazionale Superiore di Studi Avanzati, Trieste
}%
\affiliation{%
 CNR-IOM---Istituto Officina Materiali, DEMOCRITOS SISSA unit, Trieste
}%
\author{Paolo Pegolo\,\orcidlink{0000-0003-1491-8229}}\email{ppegolo@sissa.it}
\affiliation{%
 SISSA---Scuola Internazionale Superiore di Studi Avanzati, Trieste
}
\date{\today}

\begin{abstract}
The time-honored Allen-Feldman theory of heat transport in glasses is generally assumed to predict a finite value for the thermal conductivity, even if it neglects the anharmonic broadening of vibrational normal modes. We demonstrate that the harmonic approximation predicts that the bulk lattice thermal conductivity of harmonic solids inevitably diverges at any temperature, irrespective of configurational disorder, and that its ability to represent the heat-transport properties observed experimentally in most glasses is implicitly due to finite-size effects. Our theoretical analysis is thoroughly benchmarked against careful numerical simulations. Our findings thus reveal that a proper account of anharmonic effects is indispensable to predict a finite value for the bulk thermal conductivity in any solid material, be it crystalline or glassy. 
\end{abstract}

\maketitle

In a series of highly influential papers spanning the nineties~\cite{allen1989thermal,allen1993thermal,feldman1993thermal,allen1999diffusons,feldman1999numerical}, Allen and Feldman (AF) laid the ground for a harmonic theory of heat transport in glasses, which is still considered a landmark in the field. In a nutshell, the AF theory stipulates that disorder alone is able to bring down the heat conductivity from the infinite value it would have in a harmonic crystal to the finite value that is observed in a glass, without resorting to any anharmonic effects~\cite{allen1989thermal, feldman1993thermal, kittel1949}. 
This claim notwithstanding, it was soon realized that an infrared singularity inevitably affects any harmonic theory of heat transport in glasses~\cite{feldman1993thermal}. Indeed, a continuous Debye model for the low-frequency/long-wavelength vibrations combined with a standard Rayleigh quartic ($\propto \omega^4$) damping of sound waves expected from harmonic disorder ~\cite{allen1998evolution,izzo2018mixing,izzo2020rayleigh,schirmacher2006thermal} would result in a divergence of the thermal conductivity at all temperatures~\cite{feldman1993thermal, feldman1999numerical,schirmacher2006thermal,larkin2014thermal}. The impact of such singularity on the overall consistency of the AF theory and on the validity of the computations based on it has long been overlooked. On the theoretical side, it was proposed that the same quantum-tunneling effects alleged to determine the low-temperature plateau in the conductivity-temperature curve~\cite{jackle1972ultrasonic,phillips1987two,sheng1991heat,buchenau1992, feldman1993thermal,leggett2013tunneling,lubchenko2003} could also regularize the infrared singularity at high temperatures~\cite{allen1993thermal}. In the numerical applications of the AF theory, the singularity is regularized without relying on any such tunneling effects. Instead, the regularization takes advantage of the finite size of any glass model used in practice, which naturally introduces a low-frequency cutoff,  ${\omega_\mathrm{min} \sim \frac{2\pi c}{L}}$, $c$ being the sound velocity and $L$ the linear size of the system. In addition, the discrete spectrum resulting from this finite size requires an \emph{ad hoc} broadening of the vibrational lines to be dealt with, which also has a regularizing effect.

Building on our previous work on extrapolating bulk transport coefficients from finite glass models~\cite{fiorentino2023hydrodynamic}, this paper delves into the impact of the infrared singularity in the harmonic theory of heat conduction. We demonstrate that, by treating anharmonic effects within the quasi-harmonic Green-Kubo (QHGK) theory~\cite{isaeva2019modeling}, the singularity can be effectively regularized in the bulk limit at any finite temperature without relying on quantum-tunneling effects, nor on any arbitrary infrared cutoff. Our results shed light onto the ``unreasonable'' effectiveness that the AF theory has demonstrated over three decades, in spite of the infrared singularity that inherently affects it. On the one hand, we find that the contribution of frequencies below the infrared cutoff, $\omega < \omega_\mathrm{min}$, which diverges in the harmonic approximation, is relatively small when anharmonic effects are properly accounted for. On the other hand, we find that the commonly employed smearing procedure effectively mimics the boundary-scattering effects observed in thin-film samples. These findings highlight the intricate interplay between boundary and finite-size effects, on one side, and theoretical predictions, on the other, thus emphasizing the nuanced nature of the AF theory's success. Unearthing the reasons of this success provides a solid ground for advancing the theory and numerical simulation of heat transport in glassy materials.

The structure of the article is the following: first, we briefly review the AF theory and its natural extension to account for  anharmonic effects perturbatively, namely the QHGK method~\cite{isaeva2019modeling}. In both approaches the contribution of low-frequency modes to the heat conductivity can be described by a Debye model, whose parameters can be estimated from the vibrational dynamical structure factor (VDSF). Then, using the Debye model, we show that the AF prediction for the bulk thermal conductivity diverges at any temperature. This is both motivated theoretically and demonstrated numerically by an accurate finite-size scaling analysis of the thermal conductivity and VDSF of three paradigmatic glasses, amorphous silicon, silica, and silicon carbide. We then examine how this divergence is cured by either boundary-scattering or anharmonic effects and discuss the relevance of our findings to experiments performed on thin films. Finally, we present our conclusions.

\section*{Theory}

The AF expression for the heat conductivity of a glass in the harmonic approximation reads~\cite{allen1989thermal}:

\begin{align}\label{eq:kappa af}
    \kappa = \frac{\pi}{3V} \sum_{\mu\nu} C_\mu \abs{v_{\mu\nu}}^2 \delta(\omega_\mu - \omega_\nu),
\end{align}
where $\mu$ and $\nu$ enumerate normal modes, $\omega_\mu$ is the corresponding (angular) frequency, $C_\mu=\hbar\omega_\mu \left.\frac{\partial n(\omega_\mu)}{\partial T}\right|_V$ the contribution of the $\mu$-th normal mode to the isochoric heat capacity---$n(\omega)=\left[e^\frac{\hbar \omega}{k_B T}-1 \right]^{-1}$ being the temperature derivative of the Bose-Einstein occupation number, and $k_B$ the Boltzmann constant---$V$ is the volume, and $v_{\mu\nu}$ is a generalized velocity matrix. The velocity matrix, whose precise definition can be found in Refs.~\onlinecite{allen1989thermal,isaeva2019modeling}, is essentially the first real-space moment of the matrix of interatomic force constants. It is anti-Hermitean and, in a crystal, it can be chosen to be diagonal in the (Bloch) normal-mode representation, so that its diagonal elements are the group velocities of the normal modes. In a disordered system, where normal modes are necessarily real, the corresponding diagonal elements of the velocity matrix vanish, and the heat conductivity results from the coupling between (quasi-) degenerate states (see below). It must be stressed that Eq.~\ref{eq:kappa af} holds only in the thermodynamic limit, where the vibrational spectrum is continuous and the double sum actually means a double integral, while practical calculations on finite models require smearing the Dirac delta to a peaked function such as a Lorentzian, thus turning Eq.~\eqref{eq:kappa af} to~\cite{allen1989thermal, allen1993thermal}

\begin{align}\label{eq:kappa af eta}
    \kappa = \frac{1}{3V} \sum_{\mu\nu} C_\mu \abs{v_{\mu\nu}}^2 \frac{\eta}{(\omega_\mu - \omega_\nu)^2 + \eta^2},
\end{align}
where $\eta$ is the broadening width of the smeared Dirac delta. The value of $\eta$ is customarily chosen large enough to encompass several normal modes within the Lorentzian, while still remaining small enough to preserve the characteristics of a peaked function. The broadening of the delta function determines the extent to which pairs of quasi-degenerate states contribute to the heat conductivity, as strict degeneracy holds a zero probability in any finite model of a disordered system. As it will turn out, and at variance with what is commonly assumed, $\eta$ plays a crucial role in determining the value of the thermal conductivity in the AF model. In particular, its relevance becomes apparent for low-frequency vibrations within the long-wavelength regime, where normal modes gradually approach the behavior of (plane) sound waves.

In recent years, the AF approach has been generalized to incorporate a perturbative treatment of anharmonic effects, resulting in the QHGK theory~\cite{isaeva2019modeling, fiorentino2023green} or, equivalently, the Wigner Transport Equation~\cite{simoncelli2019unified, simoncelli2022wigner, caldarelli2022manybody}. The QHGK expression for the thermal conductivity of an isotropic material reads:

\begin{align}\label{eq:kappa qhgk}
    \kappa = \frac{1}{3V} \sum_{\mu\nu} C_{\mu\nu} \abs{v_{\mu\nu}}^2 \frac{\gamma_\mu+\gamma_\nu}{(\omega_\mu - \omega_\nu)^2 + (\gamma_\mu + \gamma_\nu)^2},
\end{align}
where $\gamma_\mu$ is the anharmonic linewidth of the $\mu$th normal mode, and

\begin{align}\label{eq:two-mode heat capacity}
    C_{\mu\nu}=\frac{\hbar^2\omega_\nu\omega_\mu}{T}\frac{n(\omega_\nu)-n(\omega_\mu)}{\hbar(\omega_\mu-\omega_\nu)}
\end{align}
is a generalized two-mode heat capacity. When $\omega_\mu=\omega_\nu$, Eq.~\eqref{eq:two-mode heat capacity} reduces to the modal specific heat, $C_\mu$, appearing in Eqs.~\eqref{eq:kappa af} and~\eqref{eq:kappa af eta}. The QHGK thermal conductivity, Eq.~\eqref{eq:kappa qhgk}, applies to crystalline and amorphous solids alike. For crystals, it reduces to the results of Boltzmann transport equation in the relaxation-time approximation, supplemented with inter-band effects; for glasses in the harmonic limit, the QHGK approximation reduces to the AF model~\cite{isaeva2019modeling}. Again, in practical calculations on finite systems, the AF model is restored bringing $\gamma_\mu$ from its temperature- and mode-dependent value to a temperature- and mode-independent value. The use of a constant linewidth has minimal impact on intermediate- to high-frequency vibrations---which were dubbed \emph{diffusons} and \emph{locons} by AF, due to their localization in real space~\cite{allen1999diffusons}---since, for these modes, anharmonic lifetimes are usually small and the density of states is large and slowly varying. In fact, in the AF model the diffuson contribution to the thermal conductivity is weakly dependent on $\eta$, reaching convergence once the smearing is of the order of the average normal-mode frequency spacing. On the contrary, low-frequency vibrations, referred to as \emph{propagons} by AF due to their ability to propagate like sound waves~\cite{allen1999diffusons}, display a distinct behavior. For these excitations, the vibrational density of states (VDOS) decreases quadratically as the frequency approaches zero, and the anharmonic lifetimes diverge due to the lack of vibrational decay channels. Consequently, the finite, constant linewidth introduced by smearing the Dirac delta function could possibly result in a nonphysical contribution to the heat conductivity.

In order to address the propagon contribution to the heat conductivity, it is expedient to define the vibrational dynamical structure factor. As mentioned above, propagons, diffusons, and locons differ by the degree of localization they feature. This can be observed in the VDSF that, for a harmonic system, is defined as~\cite{feldman1999numerical}:

\begin{align}\label{eq:VDSF0}
    S_b^\circ(\omega,\mathbf{Q}) =\sum_{\nu}\delta(\omega-\omega_\nu)| \langle \nu| \mathbf{Q},b\rangle|^2,
\end{align}
where $\langle \nu| \mathbf{Q},b \rangle$ denotes the projection of the $\nu$ normal mode over a sound (plane) wave vibration of wavevector $\mathbf{Q}$ and polarization $b$ ($b=L,T$ for longitudinal and transverse branches, respectively)~\cite{fiorentino2023hydrodynamic}. Anharmonic effects in the VDSF can be accounted for by smearing the delta function in Eq.~\eqref{eq:VDSF0} to a Lorentzian function:

\begin{align}\label{eq:dsf anharmonic}
    S_b(\omega,\mathbf{Q})=\frac{1}{\pi}\sum_{\nu}\frac{\gamma_\nu}{\gamma_\nu^2+(\omega-\omega_\nu)^2}|\langle \nu| \mathbf{Q},b\rangle|^2. 
\end{align}
The low-frequency, small-wavevector, portion of each branch of the VDSF features an almost linear dispersion typical of acoustic waves, $\omega_{Qb}=c_b Q$, where $c_{L/T}$ are the longitudinal/transverse speeds of sound~\cite{fiorentino2023hydrodynamic}. In other words, $S_b(\mathbf{Q}, \omega)$ is a peaked function centered at $c_b Q$ which can be faithfully represented by a single Lorentzian profile,

\begin{align}\label{eq:dsf anharmonic-2}
    S_b(\omega,\mathbf{Q}) \approx \frac{\alpha_b(\mathbf Q)}{\pi} \frac{\Gamma_b(\mathbf Q)}{(\omega-c_b Q)^2+\Gamma_b(\mathbf Q)^2},
\end{align}
allowing one to evaluate the speed of sound as well as the wavevector dependence of the sound damping coefficients, $\Gamma_{b}(\mathbf{Q})$, accounting for both disorder and anharmonic effects on the same footing. The $\mathbf{Q}$-dependent function $\alpha_b(\mathbf{Q})$ is a global prefactor that scales the Lorentzian. 

For any given polarization, in an isotropic medium the damping coefficient can only depend on the magnitude of the wavevector, yielding $\Gamma_b({Q})$. Propagons are identified as those low-frequency/long-wavelength normal modes that contribute to the VDSF in the linear-dispersion regime. The increasing broadening of the dispersion identifies a cutoff frequency for propagons, $\omega_{P}$, often referred to as the Ioffe-Regel limit~\cite{ioffe1960progress,allen1999diffusons}. According to Eq.~\eqref{eq:dsf anharmonic-2}, below this limit vibrational modes can be approximately described by damped (plane) sound waves, characterized by the group velocities $c_{b}$ and decay times ${\tau_{b}(Q)=[2 \Gamma_{b}(Q)]^{-1}}$. Consequently, the propagon contribution to the heat conductivity can be cast into the form~\cite{fiorentino2023hydrodynamic}:
 
\begin{align}\label{eq:kappa propagons and diffusons}
    \kappa_P = \frac{1}{3V} \sum_{\mathbf{Q} b}^{c_b Q < \omega_P} g_b C(c_b Q)c_b^2\tau_{b}(Q),
\end{align}
where $\omega_P$ is the propagons' cutoff frequency, and $g_b$ is the degeneracy of the propagon branch: $g_L=1$ and $g_T=2$. In the bulk limit, when the size of the system is brought to infinity, the discrete sum over states turns into an integral through the definition of a density of states; the propagon contribution to the thermal conductivity thus takes a form reminiscent of the kinetic theory of gases~\cite{fiorentino2023hydrodynamic}:

\begin{align}\label{eq:kappa propagons}
    \kappa_P = \sum_b \frac{c_b^2}{3} \int_0^{\omega_P} C(\omega) \rho_b(\omega) \frac{1}{2\Gamma_b(\omega/c_b)} \dd \omega, 
\end{align}
where $\rho_{b}=\frac{g_{b}\omega^2}{2\pi^2c_{b}^3}$ are the $L/T$ Debye's density of states per unit volume. Eq.~\eqref{eq:kappa propagons}, which applies to both crystals and glasses, is the infinite-size limit of the propagon contribution to both Eqs.~\eqref{eq:kappa qhgk} and~\eqref{eq:kappa af}, the difference lying in whether or not $\Gamma_{b}(\omega/c_{b})$ includes anharmonic effects. Formulas such as Eq.~\eqref{eq:kappa propagons}, where hydrodynamic arguments are used to extrapolate QHGK results to the infinite-size limit~\cite{fiorentino2023hydrodynamic}, will be referred to as \emph{hydrodynamic QHGK formulas.}

In general, for low enough frequencies, one has~\cite{griffin1968brillouin,baggioli2022theory,fiorentino2023hydrodynamic}: 

\begin{align}\label{eq:GammaAB}
    \Gamma_b(\omega/c_b) \approx A_b \omega^2 + B_b \omega^4.
\end{align} 
In the harmonic approximation, $A_b=0$ and the leading order in the frequency dependence of the sound damping coefficient is quartic, $\Gamma_{b} \sim \omega^4$, due to incoherent Rayleigh scattering from elastic fluctuations of the medium~\cite{allen1998evolution}. This behavior, which is confirmed by experiments~\cite{baldi2011elastic}, can be understood through random media theory on a continuous model~\cite{izzo2018mixing,izzo2020rayleigh}, or from a microscopic perspective via harmonic perturbation theory, such as in the case of crystals with mass disorder~\cite{garg2011role,mahan2019effect} or random spring constants~\cite{allen1998evolution}. 
% By plugging the Rayleigh quartic dependence of the sound damping coefficients into 
When $A_b=0$ in Eq.~\eqref{eq:kappa propagons}, one easily sees that the propagon contribution to the heat conductivity diverges at all temperatures. On the other hand, the inclusion of anharmonic contributions ensures a quadratic dependence of $\Gamma_{b}$ on frequency, resulting in a finite thermal conductivity whenever $A_b \ne 0$~\cite{griffin1968brillouin, fiorentino2023hydrodynamic}. 
We conclude that disorder alone is insufficient to guarantee a finite bulk thermal conductivity in glasses.

\begin{figure}[t]
    \centering
    \includegraphics{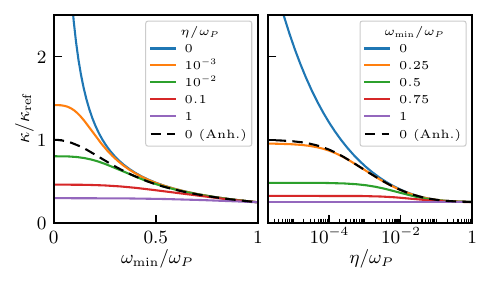}
    \caption{Heat conductivity of glasses in the harmonic approximation (see Eq.~\eqref{eq:kappa p debye}, numerical values appropriate for $\asi$). Solid lines in color are computed in the harmonic approximation, while the black, dashed, lines refer to calculations also accounting for anharmonic effects. Left panel: $\kappa$ as a function of the infrared cutoff for different values of the smearing width. Right panel: $\kappa$ as a function of the smearing width for different values of the infrared cutoff (note the logarithmic scale in the $x$-axis). Frequencies are reported in units of the propagon cutoff frequency, $\omega_P$~\cite{fiorentino2023hydrodynamic}. The thermal conductivity is reported in units of the bulk thermal conductivity in presence of anharmonic effects,  $\kappa_{\mathrm{ref}}$.}
    \label{fig:kappa theory}
\end{figure}

How come practical calculations employing the AF model yield finite values of $\kappa$ which compare fairly well with experimental results? The answer ultimately hinges on the fact that calculations are necessarily performed on finite glass models. This has two main consequences. The first is that the finite size, $L$, naturally introduces an infrared cutoff to $\kappa_P$, $\omega_\mathrm{min} \sim 2 \pi c / L$, which makes it finite. Notably, the infrared contribution to $\kappa_P$, which is divergent in the harmonic approximation, turns out to be typically small in most cases, when anharmonic effects are adequately considered as detailed below. The second consequence is that the finite number of normal modes requires the smearing of their individual contributions to Eq.~\eqref{eq:kappa af}. This leads to Eq.~\eqref{eq:kappa af eta}, wherein anharmonic linewidths are substituted with a (rather unphysical) mode-independent broadening. Therefore, the net effect of a finite calculation is that a contribution to $\kappa$---the one associated with frequencies below $\omega_\mathrm{min}$---is completely neglected, while all that remains is affected by the choice of the constant damping due to the broadening width, $\eta$. Crucially, this broadening plays a significant role even in the bulk limit, as the VDSF linewidth results to be the sum of the harmonic contribution, proportional to $\omega^4$, and the constant broadening due to the smearing~\cite{fiorentino2023hydrodynamic}. Thus, the smearing width enters the Debye expression of the harmonic thermal conductivity as

\begin{align}\label{eq:kappa p debye}
    \kappa_P = \sum_b \frac{g_b}{6 \pi^2 c_b} \int_{\omega_{\mathrm{min}}}^{\omega_P} C(\omega) \frac{\omega^2}{2 B_b \omega^4 + \eta}  \dd \omega.
\end{align}
The integral in Eq.~\eqref{eq:kappa p debye} converges for any finite value of $\eta$ and/or $\omega_\mathrm{min}$. The bulk limit is restored in the $\omega_\mathrm{min} \to 0$, $\eta \to 0$ limit. Fig.~\ref{fig:kappa theory} shows the harmonic (solid lines) and anharmonic (dashed lines) thermal conductivity of a typical amorphous solid as a function of $\omega_\mathrm{min}$ and $\eta$. The propagon contribution is obtained from Eq.~\eqref{eq:kappa p debye} (harmonic case) and Eq.~\eqref{eq:kappa propagons} with $\Gamma$ given by Eq.~\eqref{eq:GammaAB} (anharmonic). The diffuson contribution, $\kappa_D$, (which does not depend on $\omega_\mathrm{min}$) is essentially independent of $\eta$, and it thus adds the same constant shift to each line. The left panel of Fig.~\ref{fig:kappa theory} shows $\kappa$ as a function of the infrared cutoff for different values of the smearing width. When $\eta=0$, $\kappa$ diverges in the $\omega_\mathrm{min} \to 0$ limit. Vice versa, the right panel shows $\kappa$ as a function of the smearing width for different values of the infrared cutoff. Again, when $\omega_\mathrm{min}=0$, $\kappa$ diverges in the $\eta \to 0$ limit.

We must now stress that most of the experimental literature on heat transport in glasses from the nineties concerns micrometer-thick films~\cite{cahill1994thermal}, rather than samples of macroscopic size. Formally, boundary effects such as those involved in thin-film experiments would enter the expression of the thermal conductivity the same way as the AF smearing width in Eq.~\eqref{eq:kappa p debye}. For a thin-film sample, the thermal conductivity can in fact be described by an equation similar to Eq.~\eqref{eq:kappa p debye}, where $\eta$ is replaced a constant boundary-scattering contribution to the linewidth, of the form $\eta_{BS}\sim c/d$, $d$ being the film thickness~\cite{matthiessen1862influence,larkin2014thermal,cahill1994thermal,ziman2001electrons}. As a consequence, the bulk limit of the AF model with fixed $\eta$ yields the same thermal conductivity of a thin film rather than that of an infinite system. In the harmonic approximation, the former remains finite, while the latter diverges. This might have unintentionally contributed to the misconception that the heat conductivity of bulk glasses can be fully explained in terms of disorder effects alone, neglecting anharmonic interactions. These interactions dampen low-frequency vibrations and regularize the heat conductivity at all finite temperatures. In many cases, this regularization renders the infrared contribution to $\kappa$ almost negligible in the bulk limit compared to that of diffusons. Essentially, anharmonic interactions substitute a divergent quantity (the bulk thermal conductivity of propagons in the harmonic approximation) with a finite quantity that can be mimicked by a finite-size effect in calculations on finite systems. In conclusion, the presence of anharmonic interactions is vital for regularizing the behavior of $\kappa$ in a macroscopic system, even in the case of disordered materials~\cite{fiorentino2023hydrodynamic}.

\section*{Numerical experiments}

In order to substantiate our arguments, we have performed a number of numerical experiments on three glasses featuring different convergence properties to the bulk limit~\cite{fiorentino2023hydrodynamic}: amorphous silicon ($\asi$), silica ($\asio$), and silicon carbide ($\asic$). The technical details of our simulations are reported in the \emph{Methods} section.

\begin{figure}
    \centering
    \includegraphics{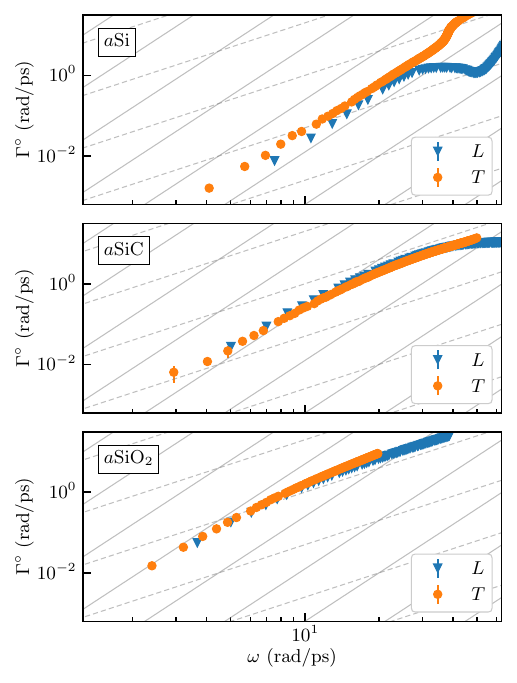}
    \caption{Sound damping coefficients as estimated in the harmonic approximation for $\asi$, $\asic$, and $\asio$. The estimate is obtained by fitting the harmonic vibrational dynamical structure factor, Eq.~\eqref{eq:VDSF0}, with its sound-wave form, Eq.~\eqref{eq:dsf anharmonic-2}, and expressing the linewidth as a function of $\omega$, as in Eq.~\eqref{eq:kappa propagons}. The sizes of the samples are respectively $13824$, $97336$ and $139968$ atoms. The estimated errors are smaller than the size of the symbols. The dashed and continuous gray lines indicate respectively the $\omega^2$ and $\omega^4$ scaling. Note the logarithmic scale on both axes.
} \label{fig:Gamma_arm}
\end{figure}

\subsection*{
Low-frequency behavior of the sound damping coefficients
}

The quartic frequency dependence of the sound damping coefficients in harmonic glasses can be understood perturbatively in terms of the scattering of acoustic waves in a homogeneous medium with small, random, independent local fluctuations of the elastic constants~\cite{allen1998evolution}. In Fig.~\ref{fig:Gamma_arm} we report the dependence of the attenuation coefficients in the harmonic approximation on frequency for the three materials considered in this work. Both $\asic$ and $\asio$ exhibit an $\omega^4\to\omega^2$ crossover, $\omega_{XO}$, respectively around $2$ and $1\,\mathrm{THz}$ ($\omega_{XO}\approx 12$ and $\omega_{XO}\approx 6\,\mathrm{rad/ps}$), in agreement with theoretical models which explain it in terms of the mixing between longitudinal and transverse modes, due to the broadening of the linear dispersion induced by disorder~\cite{izzo2018mixing}. This behavior is also in agreement with experiments, which find a first---temperature-dependent---crossover at very low frequency between an $\omega^2$ regime, determined by anharmonic effects, and the $\omega^4$ regime where disorder dominates~\cite{tomaras2010anharmonic}, followed a by second---temperature-independent---crossover from $\omega^4$ to $\omega^2$, due to the longitudinal-transverse mixing mentioned above~\cite{baldi2011elastic}. No such crossover is observed in $\asi$. As the minimum frequency compatible with a given finite glass models scales as the inverse size, rather large simulation cells are required to discriminate the crossover and evaluate the corresponding coefficients. For materials such as $\asio$, where the crossover occurs at relatively low frequencies, it is essential to have systems with several tens of thousands of atoms. In practice, standard lattice-dynamical techniques based on matrix diagonalization are unsuitable to deal with such large systems, and the VDSF can be best computed directly in these cases using Haydock's recursion method~\cite{haydock1980recursive,vast2000effects} based on the Lanczos algorithm~\cite{golub2013matrix} (see the \emph{Methods} section). We are thus able to compute the VDSF for systems comprising up to hundred-thousand atoms, an order of magnitude larger than those computed in our earlier work employing direct diagonalization~\cite{fiorentino2023hydrodynamic}.

It must be noted that the existence of the ${\omega^4\to\omega^2}$ crossover may yield misleading results in the computation of the thermal conductivity. This issue is particularly relevant for $\asio$, a material often depicted as highly disordered, whose heat conductivity would have very small finite-size effects, and reaching a well-converged value with models of a few thousand atoms~\cite{larkin2014thermal}. 
Actually, since its crossover frequency is $\omega_{XO} \sim 7.5,\mathrm{rad/ps}$ for both polarizations~\cite{baldi2011elastic}, to evaluate the bulk limit of the thermal conductivity of this system, one would need to employ samples whose linear size exceeds the wavelength of the corresponding sound wave. The wavelength for the longitudinal sound wave, $\lambda_{XO}^{L}$, is given by $\frac{\omega_{XO}}{2\pi c_L}\sim 60\,\mathrm{\AA}$, while the wavelength for the transverse sound wave, $\lambda_{XO}^T$, is given by $\frac{\omega_{XO}}{2\pi c_T}\sim \mathrm{35}\,\mathrm{\AA}$. Thus, the sample size should be greater than $\lambda_{XO}^{L}$, which means it should contain $\gtrsim 14000$ atoms.
Therefore, if one were to study the AF thermal conductivity of a glass with a finite model of linear size smaller than $\lambda_{XO}^{L}$, one would only sample the propagon contribution above the crossover, thus squarely missing the quartic low-frequency dependence of the sound damping coefficient that determines the divergence of the heat conductivity in the harmonic approximation. 

\begin{figure}[htb]
    \centering
    \includegraphics{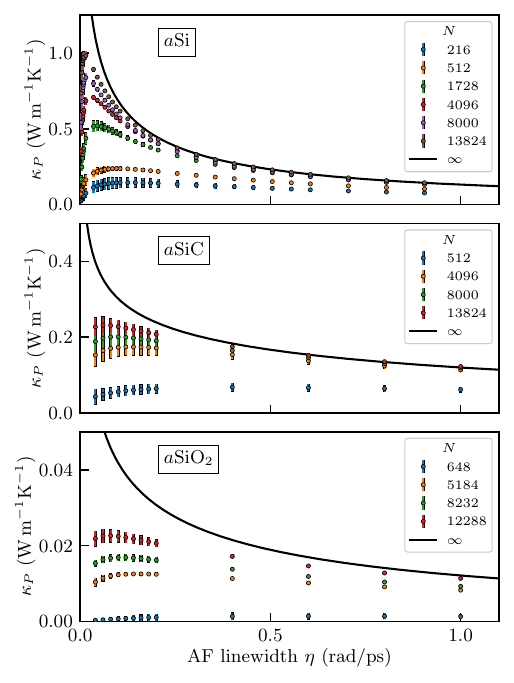}
    \caption{Thermal conductivity of propagons for models of different sizes of $\asi$, $\asic$, and $\asio$, computed with the AF model, as a function of the AF linewidth, $\eta$. Calculations are made at a temperature of $500\,\mathrm{K}$. The black, solid, line is the infinite-size analytical result, Eq.\eqref{eq:kappa p debye}.  The cutoff angular frequency for propagons is set to $\omega_P/2\pi=3\,\mathrm{THz}$, $\omega_P/2\pi=3\,\mathrm{THz}$, and $\omega_P/2\pi=1.2\,\mathrm{THz}$, for the three materials, respectively.}
    \label{fig:kappa af propagons}
\end{figure}

\subsection*{AF thermal conductivity}

In order to demonstrate how the $\omega^4$ dependence of the harmonic sound damping coefficients affects the bulk limit of the AF heat conductivity, we computed the propagon contribution to the AF conductivity in $\asi$, $\asio$, and $\asic$ over a range of values of the  smearing parameter, $\eta$, and for finite models of progressively larger sizes.  We compared these results with the analytical model provided by Eq.~\eqref{eq:kappa p debye}, $\kappa_P(T, \eta)$, whose parameters are estimated from the harmonic VDSFs. The results for $\asi$ at $500\,\mathrm{K}$ are shown in the upper panel of Fig.~\ref{fig:kappa af propagons}. As the size of the model increases, the AF data approach the analytical benchmark.  The convergence is achieved at larger sizes as the value of $\eta$ decreases. In fact, calculations on a finite system with small $\eta$ are meaningless: when the average frequency spacing of propagons is larger than the AF smearing, the Lorentzian functions in Eq.~\eqref{eq:kappa af eta} become so sharp that the corresponding effective VDOS features unphysical gaps that result in a spurious reduction of the thermal conductivity.

From the central and lower panels of Fig.~\ref{fig:kappa af propagons} similar conclusions can be drawn for $\asic$ and $\asio$, respectively.  Unlike $\asi$, both materials present the aforementioned $\omega^4\to\omega^2$ crossover around  $\omega_{XO}=12\,\mathrm{rad/ps}$ and $6\,\mathrm{rad/ps}$, respectively. This requires a different functional form for the harmonic linewidth, able to capture the crossover, such as the one proposed in Ref.~\cite{baldi2011elastic}:
\begin{align}\label{eq:gamma crossover}
    \Gamma^\circ(\omega) = C_b \omega^2 [1 +(\omega_{XO}^b/\omega)^{2\delta}]^{-1/\delta},
\end{align}
where $C_b$ is a constant, $\omega_{XO}^b$ is the polarization-dependent crossover angular frequency, and $\delta=1.5$ determines the sharpness of the transition. We then compared the AF data with the analytical model provided by Eq.~\eqref{eq:kappa propagons} with the linewidth computed with Eq.~\eqref{eq:gamma crossover}. Like for $\asi$, when $\eta$ is large, for both $\asic$ and $\asio$ the AF results converge in size to the analytical ones, and the convergence is reached at larger sizes as $\eta$ diminishes. In the $\eta\to0$ limit, the analytical model diverges due to the Rayleigh ($\propto\omega^4$) scattering term in the harmonic linewidth.

\begin{figure}[htb]
    \centering
    \includegraphics{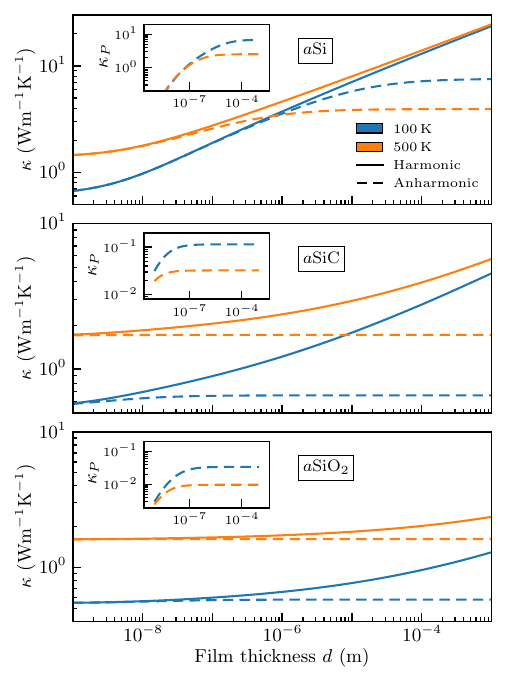}
    \caption{Thermal conductivity of $\asi$, $\asic$ and $\asio$ at $100\,\mathrm{K}$ and $500\,\mathrm{K}$, as a function of the film thickness. The solid line shows the harmonic results, the dashed line displays the thermal conductivity when anharmonicity is considered. In the inset we display the propagon thermal conductivity when anharmonicity is considered. Note the logarithmic scale on both axes.}
    \label{fig:kappa film}
\end{figure}

\subsection*{Discussion}

\begin{figure}[htb]
    \centering 
    \includegraphics{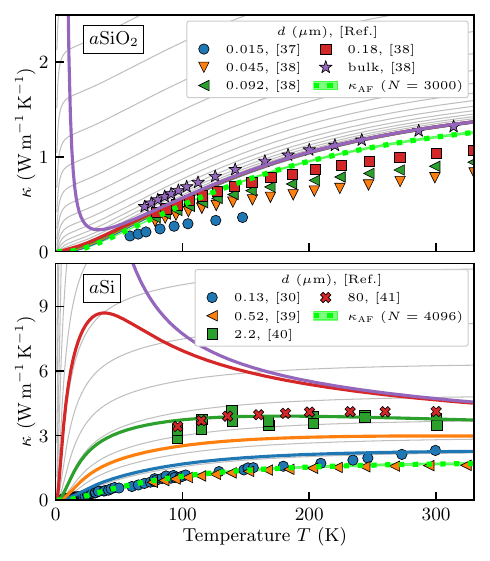}
    \caption{Thermal conductivity of $\asio$ and $\asi$ as functions of temperature. Markers are experimental data taken from Refs.~\onlinecite{swartz1987thermal, lee1997heat} ($\asio$) and Refs.~\onlinecite{zink2006thermal,cahill1994thermal,yang2010anomalously,liu2009high} ($\asi$). The solid lines, color-matched to the markers, represent hydrodynamic QHGK results for films of the same thickness. Light-green, dotted lines are standard AF results on a finite system of the size indicated and obtained with a smearing parameter $\eta$ set to the average angular frequency spacing. Gray lines correspond to bulk AF results with $\kappa_P$ computed according to Eq.~\eqref{eq:kappa p debye}, where $\omega_{\mathrm{min}}=0$ and $\eta$ ranges from $0.1\,\mathrm{rad/ps}$ (bottom-most curve) to $10^{-9}\,\mathrm{rad/ps}$ (top-most curve) on a logarithmic scale.}
    \label{fig:kappa vs T comparison}
\end{figure}

As discussed above, the AF method is commonly acknowledged to effectively account for experimental measurements of the heat conductivity of amorphous solids. To gain a deeper insight into this effectiveness, in Fig.~\ref{fig:kappa film} we analyze the dependence of the extrapolated [Eq.~\eqref{eq:kappa p debye}, ${\omega_\mathrm{min}=0}$] thermal conductivity of $\asi$, $\asic$, and $\asio$ at $100\,\mathrm{K}$ and $500\,\mathrm{K}$ on the thickness of the sample, $d$. The boundary scattering adds to the linewidth of each polarization a term equal to $\eta^b_{\mathrm{BS}}=c_b/d$~\cite{matthiessen1862influence}. The insets show the propagon contribution to the thermal conductivity, $\kappa_P$, where third-order anharmonic effects are computed with the Fermi's Golden Rule and included through the Matthiessen rule~\cite{matthiessen1862influence}. In the harmonic approximation, the thermal conductivity diverges as the film thickness increases, as indicated by the solid lines. However, when anharmonicity is accounted for, the thermal conductivity converges to its bulk value at a finite thickness. The figure demonstrates that in materials such as $\asio$ and $\asic$, where propagons contribute marginally to heat transport compared to diffusons, the bulk limit is reached at nanometer scales. In our $\asi$ model, where propagons play a more significant role, the bulk limit is achieved at much larger sizes, around a hundred micrometers. It is worth noting that, at the typical thin-film sizes used experimentally, the harmonic value of $\kappa$ is not significantly different from the anharmonic one. This suggests that a harmonic model on a \emph{finite} system can provide a reasonable estimate of the thermal conductivity of a thin film even when extrapolated to $\omega_\mathrm{min}\to 0$, as long as boundary scattering is appropriately accounted for.

Fig.~\ref{fig:kappa vs T comparison} illustrates the temperature dependence of $\kappa$ for $\asio$ and $\asi$. We compare experimental measurements from the literature with our hydrodynamic QHGK results and AF calculations conducted on finite samples. The anharmonic linewidths are computed on a range of temperatures and extrapolated to get a continuous line, as described in Refs.~\onlinecite{braun2016size,klemens1951thermal}. These linewidths are then combined with the total linewidth using the Matthiessen rule~\cite{matthiessen1862influence}. For $\asio$, the QHGK results match the bulk experimental measurement~\cite{lee1997heat}. AF calculations, performed on a sample comprising 3000 atoms, also shows good agreement with both bulk-QHGK results and experimental data. This indicates that in the case of $\asio$, diffusons completely dominate the thermal conductivity, so that similar results are obtained neglecting contributions below $\omega_\mathrm{min}$ (as done in finite AF calculations) as well as considering the anharmonic damping of propagons (as in bulk QHGK calculations). However, a direct extrapolation of the AF results regularized with a finite $\eta$ yields values of $\kappa$ ranging from $\kappa_D$ to infinity, depending on the value of the smearing parameter.

In the case of $\asi$, where propagons are more important, the intriguing effectiveness of AF calculations in matching  experimental data is further questioned. For instance, a calculation using $4096$ atoms closely agrees with measurements on a ${0.52\,\mathrm{\mu m}}$-thick film~\cite{zink2006thermal}, seemingly validating the entire procedure. Again, what is actually happening is that the (diverging) contribution to $\kappa$ from $0$ to $\omega_\mathrm{min}$ is being set to zero rather than to the (finite and small) value it would have when accounting for anharmonicity. For $\asi$, the missing contribution is not as negligible as it is for $\asio$, resulting in a pronounced difference between the QHGK results and the AF calculation. 

The temperature dependence of the QHGK heat conductivity results from two competing contributions. One is that from diffusons, which is exponentially suppressed at low temperatures---due to the Bose-Einstein occupation function---and  saturates to a constant at higher temperatures. The other is the one from propagons, which diverges as $T \to 0$ for essentially the same reasons why it does so in crystals~\cite{fiorentino2023hydrodynamic}: first, the propagation of sound waves with wavelengths much larger than the atomic correlation length is relatively unaffected by disorder at leading order in $\omega$, and, second, the temperature dependence of $A_b$ in Eq.~\eqref{eq:GammaAB} causes the integral in Eq.~\eqref{eq:kappa propagons} to diverge for vanishing temperatures~\cite{braun2016size,klemens1951thermal}. 
The concavity of $\kappa(T)$ is thus determined by the relative magnitudes of these two contributions~\cite{fiorentino2023hydrodynamic}. In materials where propagons contribute marginally to the heat conductivity (such as $\asio$, upper panel of Fig.~\ref{fig:kappa vs T comparison}), the divergence of the bulk value of $\kappa_P(T)$ becomes noticeable primarily at low temperatures. The change in concavity is thus determined by the onset of diffusons. Conversely, when propagons dominate the thermal conductivity, the concavity might be entirely determined by $\kappa_P$, such as in the case of our model of $\asi$ (purple curve in the lower panel of Fig.~\ref{fig:kappa vs T comparison}). At low temperatures, the divergence is suppressed by boundary scattering effects in thin films, as illustrated in Fig.~\ref{fig:kappa vs T comparison}. 
Even in bulk systems, where no boundary scattering exists, the low-temperature divergence is suppressed by quantum tunneling between quasi-degenerate minima in the glass energy landscape, which leads to the plateau commonly observed at a few tens of kelvins in most glasses~\cite{jackle1972ultrasonic,phillips1987two,sheng1991heat,buchenau1992,leggett2013tunneling,lubchenko2003}. 

\section*{Conclusions}

The main result of this paper is the demonstration that disorder effects alone are not sufficient to bring the heat conductivity of a material from the infinite value it has in a harmonic crystal to the finite value observed in real glasses. In the harmonic approximation, the low-frequency portion of the vibrational spectrum yields an infinite contribution to the thermal conductivity reminiscent of its behavior in crystals: in the long-wavelength/low-frequency regime, sound waves propagate in glasses essentially the same way they do in crystals, to the relevant order in frequency. In fact, the lack of sound damping in harmonic crystals and its rapid decay in harmonic glasses ($\sim \omega^4$) both fail to effectively regularize the divergent conductivity. By contrast, a proper account of anharmonic effects makes the bulk thermal conductivity of glasses finite at any finite temperature. Still, with  anharmonicity alone, the thermal conductivity would diverge at zero temperature. In practice, at extremely low temperatures the residual divergence is suppressed by quantum-tunneling effects~\cite{jackle1972ultrasonic}, leading to the well-known thermal conductivity plateau at a few tens of kelvins. This plateau is believed to be due to the tunneling between quasi-degenerate low-energy minima in the glass energy landscape, responsible for the residual entropy in glasses~\cite{berthier2011,jackle1981,Debenedetti2001}. As our treatment is limited to the vibrational properties within a single such energy minimum, it obviously fails to address the low-temperature plateau. The description of these tunneling effects from first principles thus remains a major challenge in the physics of glasses to be addressed in the future. It is noteworthy that, in materials where the dominant influence of propagons on heat transport persists at temperatures higher than those at which quantum tunneling suppresses them, our analysis indicates that the bulk thermal conductivity should display a maximum at low temperature, which could potentially be detected experimentally.

\section*{Methods}

\subsection*{Computational details}

The glass samples used in our simulations were generated through a melt-and-quench procedure. Initially, a crystalline conventional cell was replicated $\ell$ times along each Cartesian direction. Molecular trajectories were then generated in different thermodynamic ensembles (see below), employing the velocity-Verlet algorithm implemented in the \texttt{LAMMPS} code~\cite{LAMMPS}. A time step of 1/1/0.5 fs was used for $\asio$/$\asic$/$\asi$. To ensure statistical robustness, all of our results were averaged over 4/4/10 independent samples. These samples were obtained by repeating the melt-and-quench procedure multiple times, each with a different random initialization. After
the equilibration, the atomic configurations were optimized so as to make atomic forces smaller than a preassigned threshold of $10^{-10}\,\mathrm{eV/\AA}$.

Normal modes and thermal conductivities are computed with the $\kappa\mathrm{ALD}o$ code~\cite{barbalinardo2020efficient} using second- and third-order interatomic force constants obtained from \texttt{LAMMPS}.

\subsubsection*{$\asio$}

$\asio$ was modeled with a Vashishta force field~\cite{vashishta1990}. The glass was modeled starting from the $\beta$-cristobalite cubic conventional $24$--atom unit cell with mass density of $2.20\,\mathrm{g/cm^3}$ replicated $\ell=18$ times along each Cartesian direction, comprising $\approx 140,000$ atoms in the simulation box. The crystal was originally melted at $7000\,\mathrm K$ and then quenched to $500\,\mathrm K$ in $10\,\mathrm{ns}$~\cite{Ercole2017, Ercolethesis}. The system was then thermalized at $500\,\mathrm K$ for $400$ ps and for $100$ more ps in the $NVE$ ensemble. The final average density of the $\asio$ samples thus obtained is $2.408\,\mathrm{g/cm^3}$ with a standard deviation across different samples of $0.002\,\mathrm{g/cm^3}$. 

\subsubsection*{$\asic$}

$\asic$ was also modeled with a Vashishta force field~\cite{vashishta2004,vashishta2007}. The starting configuration was a crystalline cubic zinc-blend structure, with 8 atoms in the unit cell, and a mass density of $3.22\,\mathrm{g/cm^3}$ repeated $23$ times along each Cartesian direction, thus comprising $\approx 97,000$ atoms in the simulation cell. Following the procedure described in Ref.~\cite{vashishta2004}, we initially heated the crystal from $300\,\mathrm{K}$ to $4000\,\mathrm{K}$ in the $NpT$ ensemble at constant null pressure and then quenched to $500\,\mathrm{K}$ in $400$ps and finally equilibrated in the $NVE$ ensemble for $80$ ps. The average density of the $\asic$ is $2.976\,\mathrm{g/cm^3}$ with a standard deviation across different samples of $0.002\,\mathrm{g/cm^3}$.

\subsubsection*{$\asi$}
$\asi$ was modeled with the Tersoff force field~\cite{tersoff1988}. The starting configuration was a diamond structure, with 8 atoms in the unit cell, with a mass density of $2.31\,\mathrm{g/cm^3}$ repeated $12$ times along each Cartesian direction, corresponding to $\approx 14,000$ atoms in the simulation cell. At variance with $\asio$ and $\asic$, in the case of $\asi$, this moderate size already allows one to observe the $\omega^4$ scaling of the harmonic linewidth. The crystal is initially melted at $6000\,\mathrm{K}$ and then quenched to $300\,\mathrm{K}$ in $22\,\mathrm{ns}$ and equilibrated in the $NVE$ ensemble for $10\,\mathrm{ns}$~\cite{csanyi2018}. The average mass density of the $\asi$ is $2.275\,\mathrm{g/cm^3}$ with a standard deviation across samples of $0.003\,\mathrm{g/cm^3}$.

\begin{figure}[htb!]
    \centering    \includegraphics{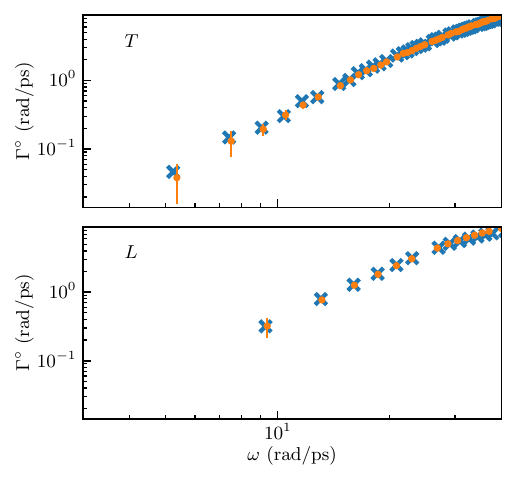}
    \caption{Sound damping coefficients as estimated in the harmonic approximation for $\asic$ obtained by fitting the harmonic vibrational dynamical structure factor, Eq.~\eqref{eq:VDSF0}, computed via Haydock's method (orange dots) and via direct diagonalization (cyan crosses). Both results are averaged over $10$ samples of $13,824$ atoms, and the error bars represent standard deviations. Upper panel, transverse modes; lower panel, longitudinal modes.}
    \label{fig:test lanczos}
\end{figure}

\subsection*{Haydock’s recursion method}
The direct computation of the harmonic VDSF is unfeasible for systems of tens of thousands of atoms because it requires the  diagonalization of the entire dynamical matrix, a procedure that scales as the cube of the number of atoms. Haydock's recursion method is an iterative procedure, based on the Lanczos orthogonalization algorithm, that allows one to estimate the VDSF as the imaginary part of a diagonal element of the vibrational Green's function of the system~\cite{haydock1980recursive,vast2000effects}.
Using this procedure, we were able to address several systems of tens of thousands of atoms, where the quartic scaling of the harmonic linewidth is appreciable, as reported in Fig.~\ref{fig:Gamma_arm}. 

We want to compute the diagonal matrix elements of the vibrational Green's function of the form:

\begin{multline}
    \lim_{\epsilon\to0} \Im\expval{\left((\omega+i\epsilon)^2 -\widebar{\mathbf{K}}\right)^{-1}}{\mathbf{Q}, b}=\\ \frac{\pi}{2\abs{\omega}}\left[S_b^0(\omega, \mathbf{Q})+S_b^0(-\omega, \mathbf{Q}) \right],
\end{multline}
where 

\begin{align}
    \widebar{\mathbf{K}}= \mathbf{M}^{-1/2}\mathbf{K}\mathbf{M}^{-1/2},
\end{align}
$\mathbf{K}$ is the matrix of interatomic force constants (i.e. the Hessian of the energy with respect to atomic displacements), and $\mathbf{M}$ is the diagonal, positive-definite, matrix of the atomic mass distribution. In a system of $N$ atoms, $\ket{\mathbf{Q}, b}$ is  a $3N$-dimensional vector whose projection onto the displacement of $I$-th atomic site in the $\alpha$-th Cartesian direction is:

\begin{align}
    \bra{I, \alpha}\ket{\mathbf{Q}, b}=\frac{1}{\sqrt{N}}\epsilon_\alpha^b(\mathbf{Q})e^{i \mathbf{Q} \cdot \mathbf{R}_I},
\end{align}
where $\epsilon^b(\mathbf{Q})$ is the polarization vector, and ${\mathbf{Q}=\frac{2\pi}{L}(n,m,l)}$, with $(n,m,l)\in \mathbb{Z}^3$, is a wavevector compatible with the enforced PBCs. The harmonic VDSF is then computed by a continued fraction expansion:

\begin{widetext}
    \begin{equation}\label{eq: haydock-method}
        \frac{\pi}{2\abs{\omega}}\left[S_b^0(\omega, \mathbf{Q})+S_b^0(-\omega, \mathbf{Q})\right]=  % \\
        \lim_{\epsilon\to 0}\Im\frac{1}{(\omega+i\epsilon)^2-a_0-\frac{\displaystyle b_1^2}{\displaystyle (\omega+i\epsilon)^2-a_1-\frac{\displaystyle b_2^2}{\displaystyle \ddots}}},
    \end{equation}
\end{widetext}
where the coefficients $\{a_0, a_1,\dots\}$ and $\{b_1, b_2, \dots\}$ are evaluated by the recursion Lanczos chain:

\begin{align}
    \begin{split}
    &\ket{\xi_{-1}}=0,\\
    &\ket{\xi_0}=\ket{\mathbf{Q}, b},\\    
    &b_n\ket{\xi_n}=\left(\widebar{\mathbf{K}}-a_{n-1}\right)\ket{\xi_{n-1}}-b_{n-1}\ket{\xi_{n-2}}, \\ 
    &a_n=\bra{\xi_n}\widebar{\mathbf{K}}\ket{\xi_n},\\
    &b_n=\bra{\xi_n}\widebar{\mathbf{K}}\ket{\xi_{n-1}}.
    \end{split}
\end{align}
This procedure drastically reduces the computational cost of the evaluation of the harmonic VDSF, going from a $\mathcal{O}\bigl ((3N)^3 \bigr )$ scaling of the exact diagonalization algorithm to the $\mathcal{O}(k(3N)^2)$ scaling, where $N$ is the number of atoms in the simulation cell and $k$ is the number of steps of the Lanczos chain. Moreover, since the matrix of the interatomic force constants is sparse, the numerical burden of Haydock's algorithm can be further reduced to a complexity $\mathcal{O}(kN)$. The procedure proves to be numerically robust, in spite of the well-known instabilities of the Lanczos tridiagonalization scheme~\cite{paige1980accuracy}, and approximately $200$ recursion steps are typically sufficient to estimate the sound damping coefficients, which we increased up 600 steps to carefully test the convergence. In order to validate the iterative algorithm we compared the harmonic attenuation coefficients fitted from the VDSF computed via direct diagonalization of the dynamical matrix and via Haydock's method as in Eq.~\eqref{eq: haydock-method}. In Fig.~\ref{fig:test lanczos} we display the sound damping coefficients computed on a model of $\asic$ of 13824 atoms, showing good agreement between the two methods. 

% \section*{Data Availability}

% The data and scripts that support the plots and relevant results within this paper are available on the Materials Cloud platform~\cite{talirz2020materials}. See DOI:XXX. 

% \section*{Code Availability}
% The codes that support the relevant results within this paper are available to the respective developers. Analysis scripts are available on GitHub~\cite{Fiorentino_hydro_glasses_2023} and on the Materials Cloud platform~\cite{talirz2020materials}. See DOI:[to be included when available].

\begin{acknowledgments}
The authors are grateful to Federico Grasselli for a critical reading of the early version of the manuscript. This work was partially supported by the European Commission through the \textsc{MaX} Centre of Excellence for supercomputing applications (grant number 101093374) and by the Italian MUR, through the PRIN project \emph{FERMAT} (grant number 2017KFY7XF) and the Italian National Centre for HPC, Big Data, and Quantum Computing (grant number CN00000013).
\end{acknowledgments}

\bibliography{main}

\end{document}